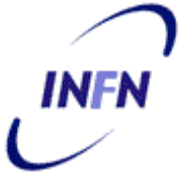

# ISTITUTO NAZIONALE DI FISICA NUCLEARE

Sezione di Trieste

# CONCEPTUAL DESIGN FOR A POLARIZED PROTON-ANTIPROTON COLLIDER FACILITY AT GSI

F. Bradamante [1], I. Koop [2], A. Otboev [2], V. Parkhomchuk [2], V. Reva [2],
P. Shatunov [2], Yu. Shatunov [2]

[1] *University of Trieste and INFN, Trieste, Italy*
[2] *Budker Institute of Nuclear Physics, Novosibirsk, Russia*

## Abstract

Two possible options of polarized proton-antiproton collider at the future HESR storage ring are considered. It is shown that the modifications of the present HESR project which are needed to arrange for the polarized proton-antiproton collisions are relatively moderate. An achievable luminosity of $5 \cdot 10^{31}$ cm$^{-2}$·sec$^{-1}$ will provide a possibility to carry out experiments in the CM energy range 10-30 GeV, a particularly interesting option for Drell-Yan physics.

## Table of Contents

# 1. Subject and objectives of this study

Subject of this study is the conceptual design of a polarized proton-antiproton scattering facility which is under consideration as a possible part of the future extension of the Facility for Antiproton and Ion Research (FAIR) at the Gesellschaft für Schwerionenforschung mbH (GSI), in Darmstadt, Germany [1]. It shall be used for a better planning of the lay-out of the proposed experiments at FAIR, and, in particular, a Spin Experiment with Anti-Protons (SEAP). Two international Collaborations (ASSIA [2] and PAX [3]) have expressed their interest for spin physics with antiprotons at FAIR To investigate the spin dependence of the partonic structure of the nucleon at the future storage ring HESR a the center-of-mass energy of at least 15 GeV, and a luminosity of proton-antiproton colliding beams above $10^{31} cm^{-2} s^{-1}$, are required.

The following topics are under consideration:
1. Proton-antiproton collisions in the HESR (15 + 15 GeV)
   - Luminosity estimations
   - Basic beam parameters
   - Storage ring layout and operation.
   - Design of spin manipulator
2. Asymmetric option of the collider (3.5 + 15 GeV)
3. Polarized protons and antiprotons production.

Main objectives of the present study are the parameters of the symmetric collider ring option and a lay-out of a suitable interaction region, taking into account intrabeam scattering, beam-beam interaction and cooling process. The relevant requirements for the design of the rings and for the beam parameters are deduced. For comparison, an asymmetric option of collider has been studied with an additional "COSY type" storage ring ($COSY$) and beam energies 3.5 + 15 GeV.

The collider is planned to be supplied by beams from the main branch of the FAIR project which is under development at GSI, with moderate modifications for polarized beams. The envisaged additional equipment for SEAP might consist of specific parts to produce the polarized proton and antiproton beams, to deliver the polarized beams into the storage ring, and to control the beam polarizations in the whole HESR energy range.

A schematic scenario of the polarized beam complex is shown in Figure 1.

The polarized protons from the source are accelerated in the linac and synchrotron SIS-18 up to an energy of 3.5 GeV and injected into HESR. Estimations of depolarizing resonances at SIS-18 have shown that one solenoidal partial Siberian snake will provide safe acceleration of the polarized protons through imperfection resonances. However, with the normalized beam emittance of 20p mm·mrad, a pulsed quadrupole is required to overcome one intrinsic spin resonance. Antiproton polarization will be achieved by multi-scattering on an internal polarized hydrogen jet target in a special AP ring, which is supplied by antiprotons from the complex of CR-RESR storage rings. The energy and the other parameters of the AP will be fixed according to the result of a test experiment of the filtering method in the AD ring at CERN. After polarization, antiprotons are injected into the *COSY* ring, equipped by e-cooler and full Siberian snake, for delivering good quality polarized beam to the collider at an energy of 3.5 GeV.

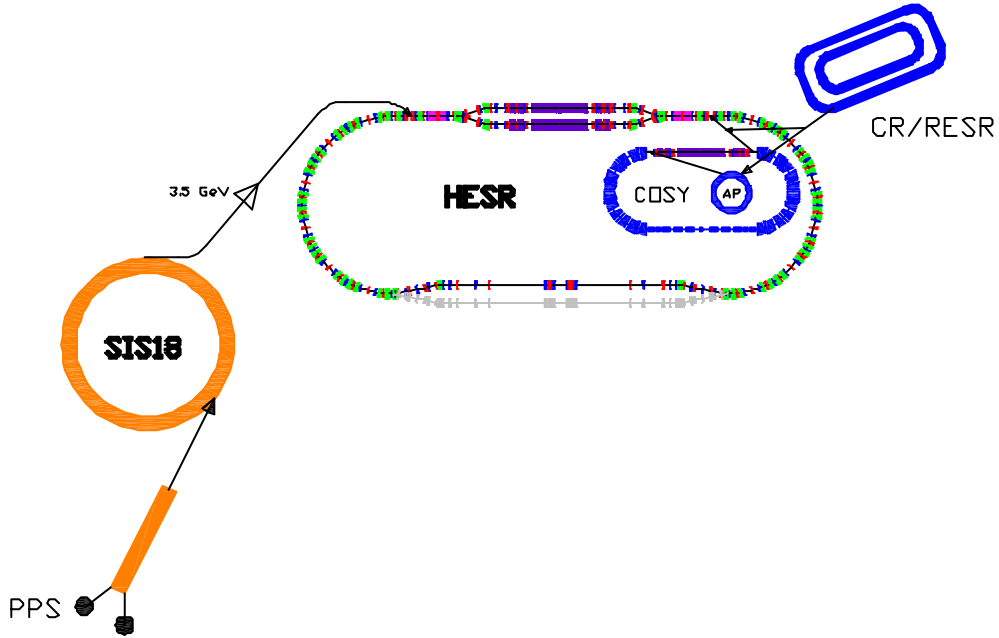

**Figure 1. Schematic layout of polarization facility**

## 2. HESR collider option

In this chapter the p-$\bar{\text{p}}$ collider, based on the use of the 15 GeV HESR storage ring, is presented. Both the proton and antiproton beams are circulating in a common vacuum chamber (see the Figure 2). We have foreseen two parallel long straight sections to accommodate (alternatively) two interaction regions. One houses the PANDA experiment (antiproton scattering on the internal target [4]) and another serves to perform a Spin Experiment with Anti-Protons (SEAP).

Antiproton and proton beams can be directed alternatively to one of two interaction regions. The distance between two bypasses equals roughly to 10 m. So, each orbit shift in x-direction (we call it "bridge") is about +5 m or -5 m. The bridge occupies about 20 m of the orbit length.

In the long straight section, opposite to the interaction region, proton and antiproton beams are electrostatically separated and directed through different long solenoids of the electron cooler. Being a part of the Siberian snake, these solenoids simultaneously serve also for the spin control. The cooler's electron beam is accelerated to 8 MeV, and will be used for cooling of both protons and antiprotons beams sequentially. To that purpose, the electron beam passes first through the cooler solenoid, which is dedicated say for cooling of the antiproton beam, and then it is bent by $180^0$ and directed into another solenoid for second beam cooling. After passing through the second solenoid the electron beam returns back to the acceleration column for deceleration and energy recovery.

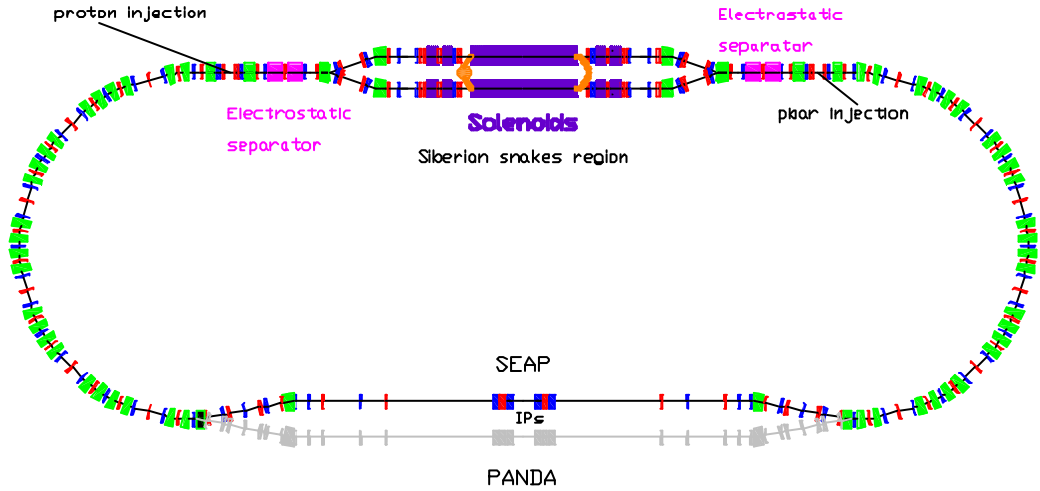

**Figure 2. Layout of the proton-antiproton collider**

## 3. Luminosity considerations

The main limitation on luminosity comes from the balance between the antiproton loss rate and accumulation rate. Luminosity cannot exceed the limit:

$$L_{max} = \frac{\dot{N}_p}{\sigma_{total}},$$

where $\dot{N}_{\bar{p}}$ is the antiproton accumulation rate (the loss rate is equal to the accumulation rate in the equilibrium state) and $\sigma_{total}$ is the total scattering cross-section.

Protons and antiprotons will be lost mainly due to their nuclear interaction. The relevant cross-section $\sigma_{total} = 40\text{mb}$ weakly depends on energy [5]. So, if the antiproton accumulation rate is limited by $\dot{N}_{\bar{p}} = 2 \cdot 10^6$ (in polarized mode), then in the whole energy range the collider luminosity is restricted by the value: $L_{max} = 5 \cdot 10^{31} \text{ cm}^{-2}\text{s}^{-1}$.

Due to the relatively high energy of protons and antiprotons, the Coulomb scattering is suppressed as compared to the nuclear interaction. For the angular acceptance of HESR $\theta_{max} = 5\text{mrad}$ we get an estimation of the Coulomb scattering cross-section:

$$s_{Coulomb} = \frac{p r_p^2}{g^2 b^4 q_{max}^2} = 12 \text{microbarn}$$

Here $r_p = 1.55 \cdot 10^{-16}$ cm is the classical electromagnetic radius of the proton. We see that it is smaller than the nuclear cross-section by three orders of magnitude.

In the following we will limit the total number of protons to the value $N_1 = 10^{12}$ distributed in $n_b = 12$ bunches. The number of antiprotons $N_2$ will be less or equal to the number of protons. At low energies the number of particles in both colliding beams could be made equal and energy independent: $N_1 = N_2 = 10^{12}$.

Let us consider various expressions for the luminosity. The luminosity of a collider with round beams can be expressed in the form:

$$L = \frac{N_1 N_2 f_0}{n_b 2p(e_1 + e_2) b_0}$$

Here $f_0$ is the revolution frequency, $e_{1,2}$ are the emittances of the first and the second beams, and $b_0$ is the beta-function value at the IP. For the flat beam geometry $b_0$ should be replaced by $\sqrt{b_{x0} b_{z0}}$.

Limitations on the beam intensity come from possible instabilities in the electron cooler. Our estimation of the instabilities threshold corresponds to a bunch population of approximately $N_b = 0.8 \cdot 10^{11}$. Another limitation comes from the space-charge effect, which gives the tune shifts:

$$\Delta n_{1,2} = \frac{N_{1,2} r_p R}{2\sqrt{2p} n_b s_s e_{1,} g(g^2 - 1)},$$

where $s_s$ is the bunch length ($s_s = b_0 = 30$ cm). The space-charge tune shifts can not exceed a certain limit, which we put $\Delta n_1 = \Delta n_2 = 0.1$, according to the international experience.

Assuming that electron cooling will squeeze proton and antiproton beams to the space charge limit, one can combine the tune shift and luminosity formulas and get the following expression for the luminosity at low energies:

$$L = \frac{N_1 \cdot \Delta n \cdot f_0 s_s g(g^2 - 1)}{\sqrt{2p} r_p R b_0}$$

This formula determines the energy dependence in the low energy range, as long as $L \leq L_{max}$. Above the threshold $T_m = 12.916$ GeV the luminosity is saturated at the level $L_{max} = 5 \cdot 10^{31}$ cm$^{-2}$s$^{-1}$. As already explained, above T=12.9 GeV the luminosity is limited by the equilibrium between the rate of particle losses and the polarized antiproton accumulation rate: $\dot{N}_2 = 2 \cdot 10^6$. Figure 3 shows the dependence of the luminosity versus energy over the entire energy range of the collider.

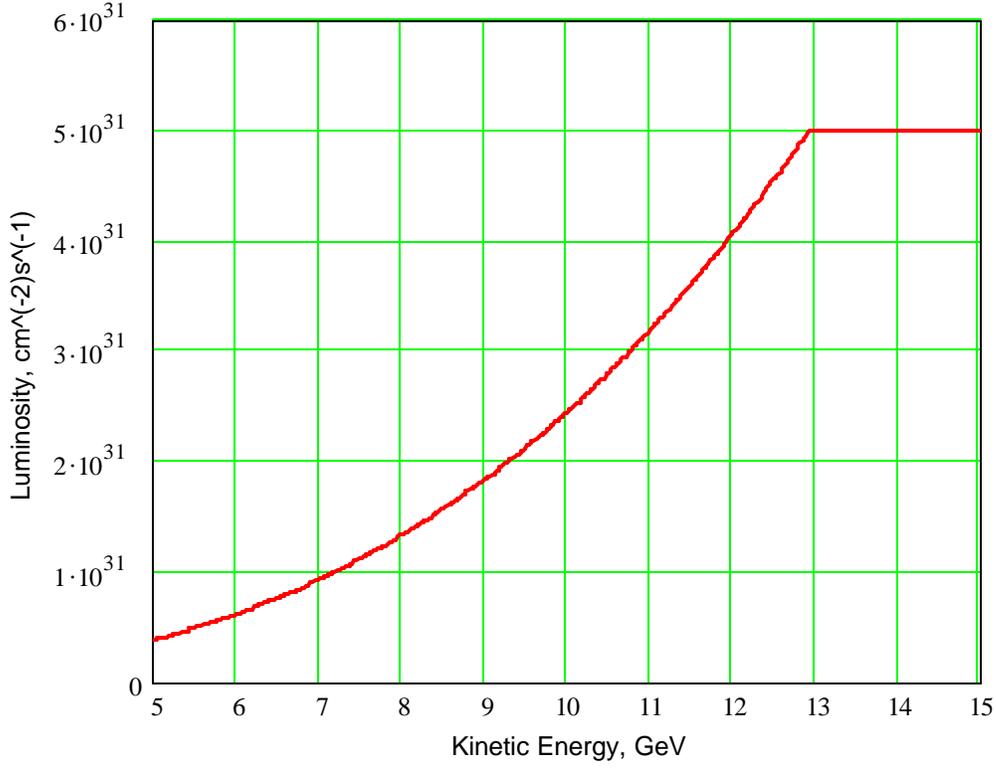

**Figure 3. Energy dependence of the luminosity**

The proton beam has a constant intensity in the whole energy range. Its emittance is governed by the inverse space-charge formula:

$$e_1 = \frac{N_1 r_p R}{2\sqrt{2\pi} n_b s_s \Delta n_1 g(g^2 - 1)}.$$

The antiproton beam has the same emittance below T=12.9 GeV ($g_m = 14.765$), but above this threshold its intensity becomes lower and its emittance follows the relation:

$$e_2 = e_1 \left( \frac{2\Delta n\, N_1 f_0 s_s g(g^2 - 1)}{L_{\max} \sqrt{2\pi}\, b_0 r_p R} - 1 \right)^{-1}.$$

Figure 4 shows the energy dependence of the proton and antiproton beams emittance, while the antiproton beam intensity as a function of energy is shown in Figure 5.

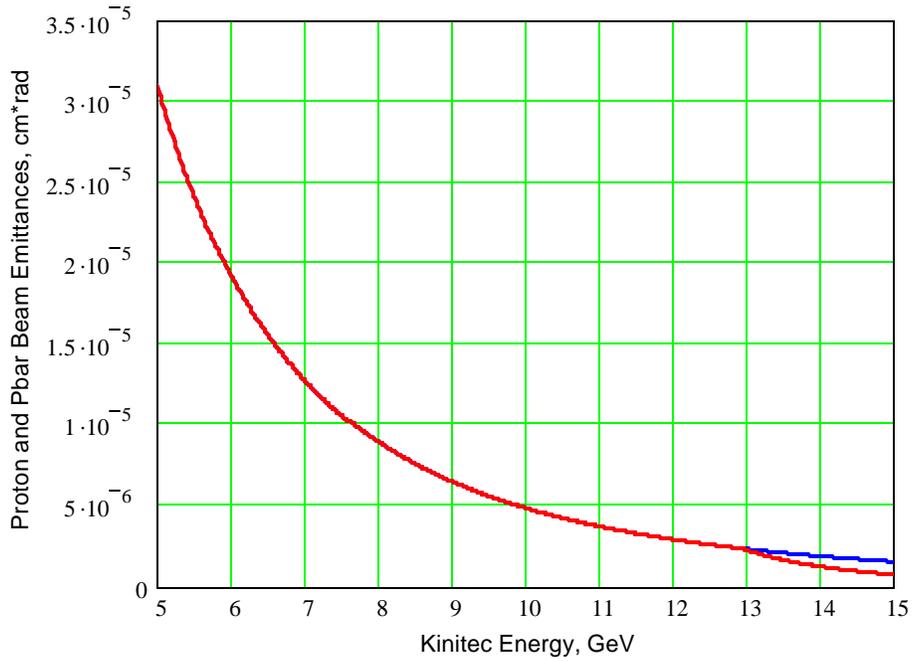

**Figure 4. The beam emittance energy dependence (protons — red-blue, antiprotons — red)**

Above $g_m$ the antiproton beam intensity is equal to $N_2 = e_2 \dfrac{2\sqrt{2p}\Delta n_1 n_b s_s g(g^2-1)}{r_p R}$.

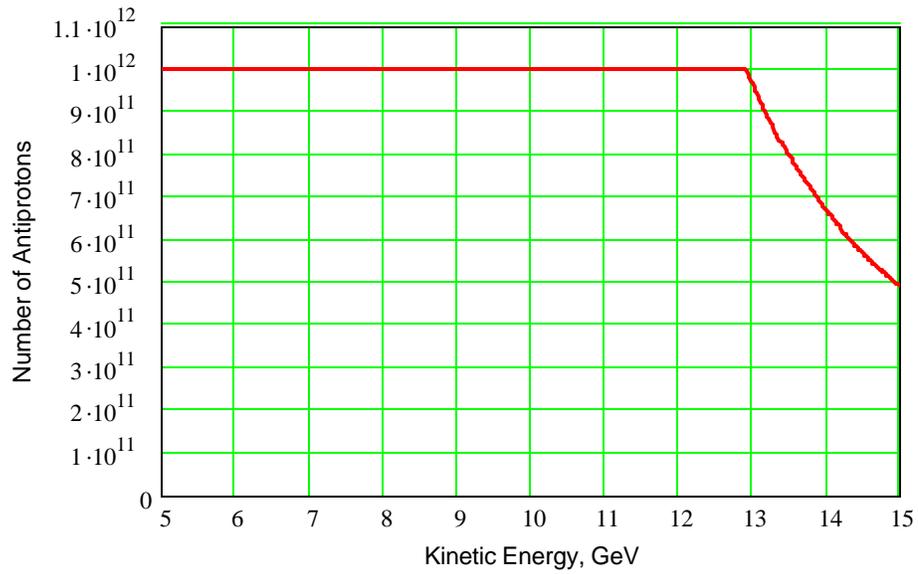

**Figure 5. Number of antiprotons vs. energy**

At the top energies $g \geq g_m$ the production rate of polarized antiprotons $\dot{N}_{\bar{p}} = 2 \cdot 10^6 \; \bar{p}/s$ is not sufficient to compensate particle losses, which are proportional to the luminosity.

An estimation of the beam-beam parameters $x_{1,2} = \dfrac{N_{2,1} \, r_p}{4\pi \, n_{2,1} \, g_{1,2} \, e_{2,1}}$ gives the values $x_1 = x_2 = 0.022$, that look quite achievable for the antiproton-proton colliding beams with the electron cooling.

## 4. Intra beam scattering and electron cooling

To achieve the required luminosity small transverse beam sizes are needed, therefore it is important to evaluate correctly the diffusion rate due to Intra Beam Scattering (IBS). In Figure 6 the IBS diffusion rates are plotted against the beam energy.

One can see that the IBS diffusion grows with energy. The damping provided by the electron cooling was estimated to be the same as that based on the proposal of Ref. 6. The cooling time at the top energy is about 10 sec (see Fig. 7). It means that even at the top energies the cooling force can suppress the IBS diffusion and, thus, keep the beam parameters in the equilibrium state.

Preliminary parameters of the cooling device for HESR are given in Table 1.

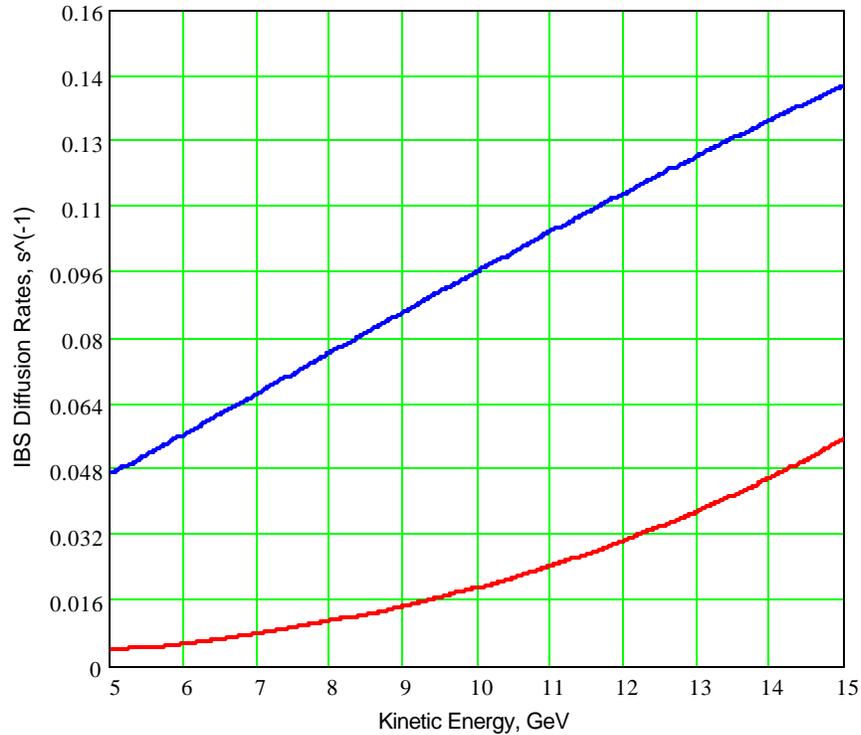

**Figure 6. IBS transverse (red) and longitudinal (blue) diffusion rates vs. energy**

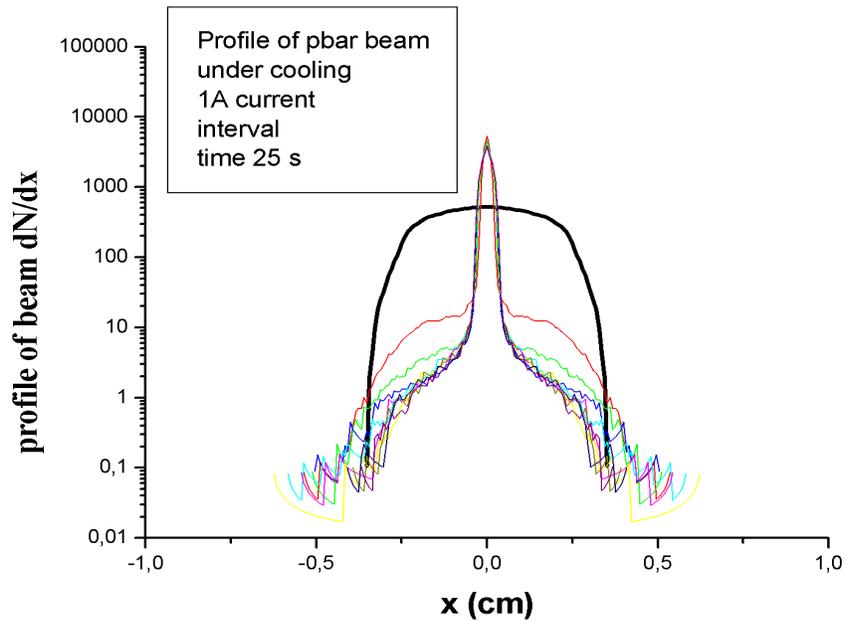

**Figure 7. Transverse beam profile at HESR under electron cooling**

**Table 1. Preliminary parameters of the electron cooler for HESR**

| **Acceleration column** | |
|---|---|
| Electron energy on the output | 0.44–7.9 MeV |
| Length | 8.0 m |
| Average electrostatic intensity along accelerator column | 0.5–10 keV/cm |
| Magnetic field | 500 G |
| Cathode diameter (beam diameter) | 2 cm |
| Height of high-voltage vessel | 13.0 m |
| Diameter of high-voltage vessel | 6.0 m |
| | |
| **Bending section** | |
| Magnetic field | 5 kG ($E_e$=1.6–7.9 MeV) |
| | 2 kG ($E_e$=0.44–1.6 MeV) |
| Bending radius | 400 cm |
| Beam diameter | 0.6 cm (5 kG) – 1.0 cm (2 kG) |
| | |
| **Cooling section** | |
| Magnetic field | 5 kG ($E_e$=1.6–7.9 MeV) |
| | 2 kG ($E_e$=0.44–1.6 MeV) |

## 5. Basic parameters of the proton-antiproton collider

Main parameters of the HESR collider option are listed in Table 2.

**Table 2. List of the proton-antiproton collider parameters**

| Collider circumference, $C$ | 681.58 | $m$ |
|---|---|---|
| Revolution frequency, $f_0$ | 0.445 | $MHz$ |
| Total number of antiprotons, $N_{\bar{p}}$ | $1 \cdot 10^{12}$ | |
| Total number of protons, $N_p$ | $1 \cdot 10^{12}$ | |
| Number of bunches per beam, $n_b$ | 12 | |
| Distance to first parasitic crossing, $L_p$ | 28.4 | $m$ |
| Proton beam emittance, $e_p$ | $1.55 \cdot 10^{-6}$ | $cm \cdot rad$ |
| Antiproton beam emittance, $e_{\bar{p}}$ | $0.465 \cdot 10^{-6}$ | $cm \cdot rad$ |
| Space charge tune shift, $\Delta n_p = \Delta n_{\bar{p}}$ | 0.1 | |
| Beam-beam parameter, $V_{x,z}$ | 0.022 | |
| Electron cooling and IBS time constants, $t_e \approx t_{IBS}$ | 10 | $s$ |
| Luminosity, $L$ | $5 \times 10^{31}$ | $cm^{-2}s^{-1}$ |

## 6. Optics of the collider

The collider's optics is designed keeping in mind the existing solutions for the HESR magnetic system [7]. It keeps the same arcs optics, and makes all matches between the different parts of the ring. New elements are: the electrostatic separators, two cooling sections and Siberian snakes in the "technical" straight section, and, on the other side of the machine, two bypasses to arrange for two different collision points for SEAP and PANDA.

Optical functions and dispersion of the whole ring are presented in the Figure 8.

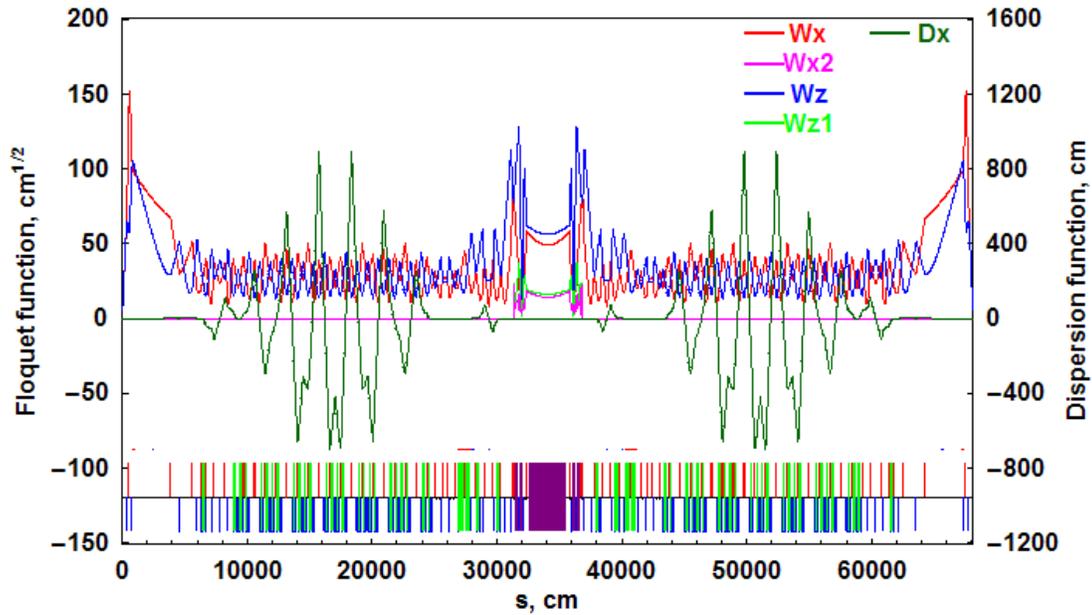

**Figure 8. Optical functions of the modified HESR**

## 6.1. Arc lattice

Each of the two arcs is made of six periods. Each period consists of three FODO cells with four $7.5^0$ bending magnets. The optical functions of the total arc are presented in Figure 9. The arc begins and ends in the middle of the D-quad. The dispersion function and the dispersion angle automatically becomes zero at the arc's entrance and exit. This is a consequence of the integer tune advance over the arc: $\Delta \nu_{x,z} = 5$.

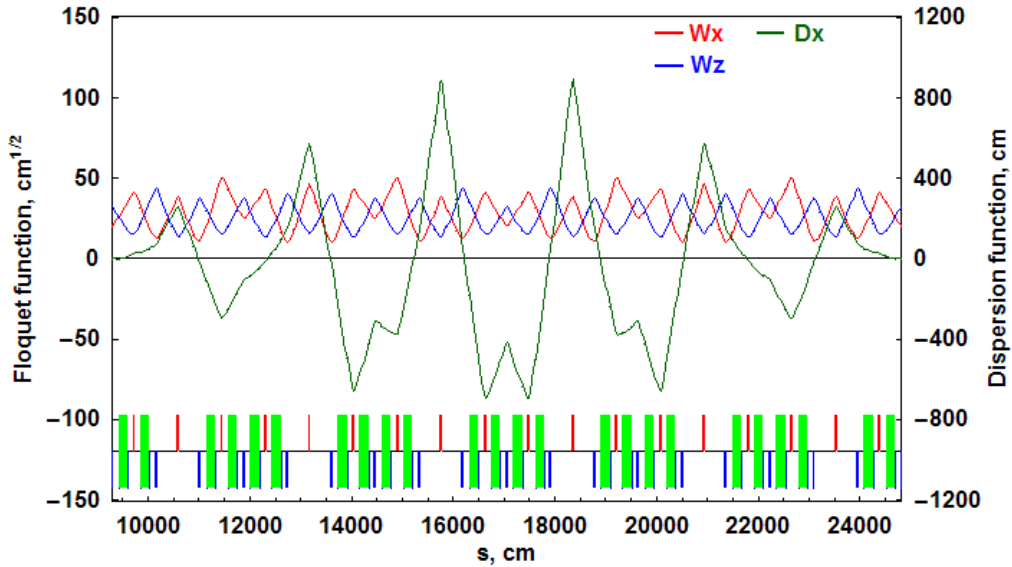

**Figure 9. The optical functions of the HESR arc**

### 6.2. Injection insertion

An injection insertion is needed to accommodate the septum magnet, the kicker and the pre-kicker. The phase advance between the centers of the pre-kicker and of the kicker is adjusted to be $180^0$. In this way, the kicker cancels the horizontal betatron oscillations excited by the pre-kicker. It is very important not to disturb the accumulated beam during the stacking process. The septum magnet is placed in the middle of the mirror-symmetric optical structure of the insertion. The optical functions of the insertion are matched to the arc's end values (see Figure 10).

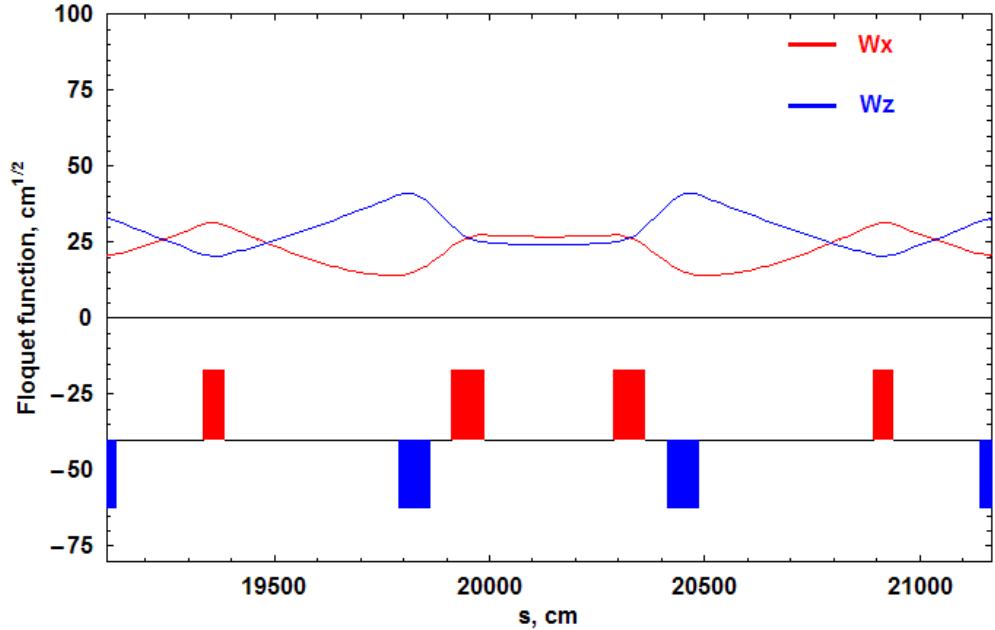

**Figure 10**. Optical functions of the mirror-symmetric insertion. The beam will be injected in the middle drift (3 m long), being bent by the septum magnet. Pre-kicker and kicker occupy the upstream and downstream drifts 4 m long.

### 6.3. Proton-antiproton electrostatic separator

The main purpose of this insertion is to split the orbits of the proton and antiproton beams to a distance of about 3.6 m, which is necessary to bend the 8 MeV electron beam backward so that it can be used to cool both beams simultaneously.

The proton and the antiproton beams need to be deflected in opposite directions, first by the electric field, which is equal to 48 kV/cm, and then by the two-window superconducting magnet, which bend the particles by $\pm 6.256^0$. The electrostatic plates (4 m long) occupy two drifts, each 5 m long. The two-window septum magnet is placed downstream, where the orbit separation becomes large enough to place there the splitting current-septa. The length of this magnet is 3 m. The vertical magnetic field at one side of the septa is B=1.825 T, while at the second window B=-1.825 T.

A compensating negative magnetic bend ($-6.402^0$) is placed further downstream. It compensates the previous electrostatic plus magnetic bends. The resulting orbit shift equals to $\pm 1.8$ m. So, the proton/antiproton cooling straight sections separation reaches 3.6 m. That is enough to make the $180^0$ electrostatic bend of the electron beam using an electric field of 45 kV/cm.

At the exit of the insertion, the optical functions $w_{x,z}$ are made equal and much bigger in values. This is done to match the ring structure with the snake insertion (see Figure 11).

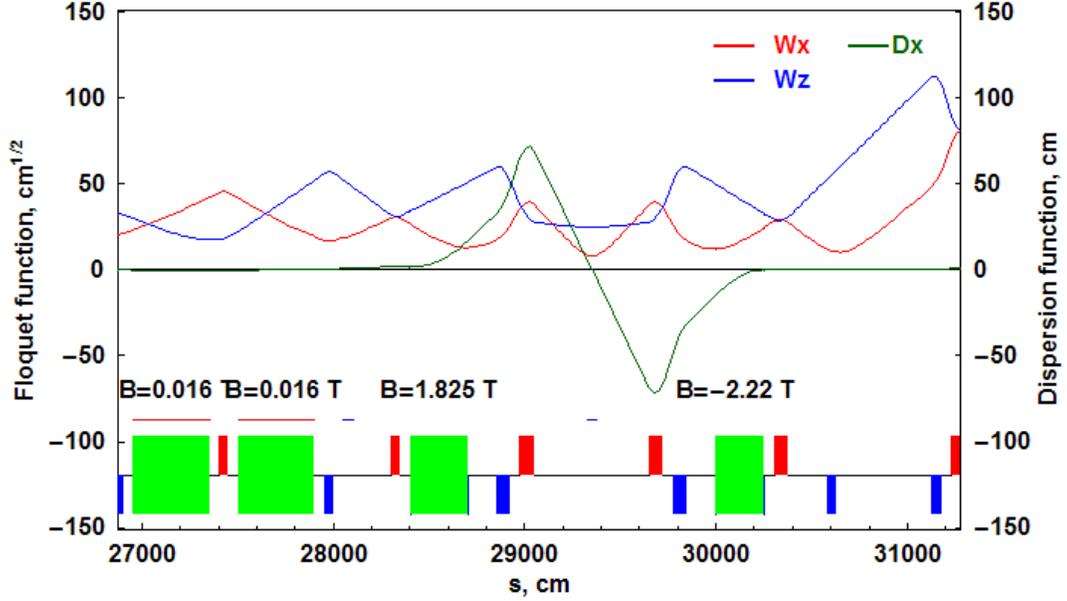

**Figure 11. Optical functions of the proton-antiproton beams separator**

### 6.4. Coolers/snakes insertion

The cooler solenoid has a length of 30 m and a magnetic field of 0.5 T. A constraint of the e-cooling device is its delicate tuning, which demands to keep the magnetic field unchanged even during the ramping. On the other hand, the field integral $\int Bds = 15\,\text{T}\cdot\text{m}$ gives a considerable spin rotation angle. We propose to increase the field integral up to the full Siberian snake value [7] ($\int Bds = 56.4\,\text{T}\cdot\text{m}$ for the top energy) by adding four additional solenoids (see the Figure 12).

To compensate the coupling and to match these solenoids (including the cooling one) with the rest of the machine optics, a number of skew quadrupoles are needed. The transformation matrix for the vertical motion is made equivalent to a drift space matrix with a length L=56 m. Initial values of the optical functions are chosen to be:

$$w_{x,z} = \sqrt{5000 + \frac{2800^2}{5000}} = 81.04\,\text{cm}^{1/2}; \quad a_{x,z} = 0.56$$

They correspond to those values, which appear with $b_{x,z}^{*} = 5000\,\text{cm}$ crossover at the center of the 5600 cm drift. The total tune advance for vertical oscillations is equal to $\Delta n_z = 1.25$, while for horizontal oscillations we have $\Delta n_x = 0.75$.

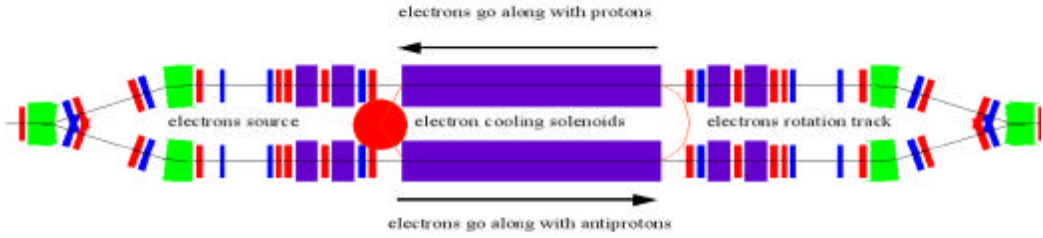

**Figure 12. Layout of the coolers/snakes insertion**

In the horizontal plane the transformation matrix is minus the matrix of a drift:
$$T_x = -T_z, \quad T_z = \begin{pmatrix} 1 & L \\ 0 & 1 \end{pmatrix},$$
where $L = 56\,\text{m}$ is the length of the whole insertion. The optical functions inside the insertion are presented in Figure 13.

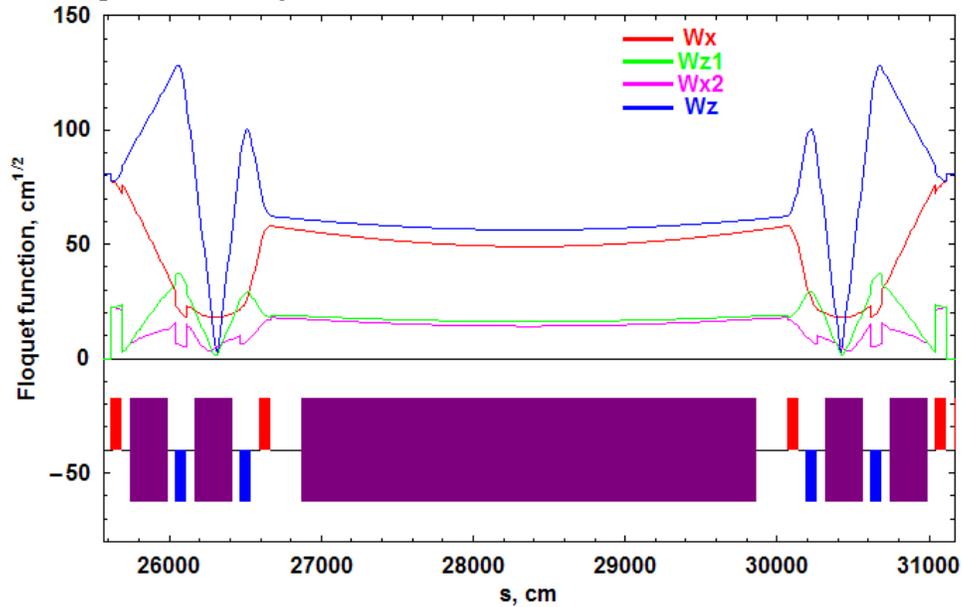

**Figure 13. Optical structure of the cooler/snake insertion**

To compensate the coupling due to the solenoids, all quadrupoles should be rotated by the angle:

$$f = \frac{1}{2}\int_0^s Bds \bigg/ Br$$

Here the integral of the longitudinal magnetic field is taken from the middle point of the insertion up to the quadrupole location azimuth "s". Starting from the cooling solenoid, the first and second quadrupoles are tilted by $\phi_1$; the third one by $\phi_2$ and the last one by $\phi_3$. Here $\phi_1$ and $\phi_2$ are dependent on the beam energy, but $\phi_3$ is *constant*, because the Siberian snake has to rotate the spin always by 180 degrees. It is necessary to remark that the right and left quadrupoles have to be rotated by opposite angles.

Figure 14 gives the fields in the Snake solenoids, and Figure 15 shows the angles $\phi_i$ versus the beam energy.

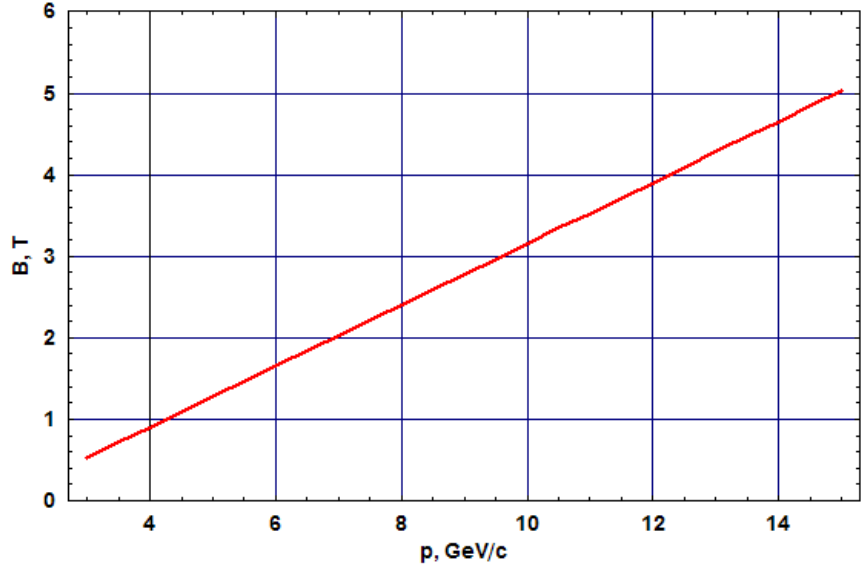

**Figure 14. Magnetic field in the snake solenoids vs. the beam energy**

In case of operation with e-cooling and without polarization (snake solenoids off) all skew-quads have to be tilted by the same angle $\phi_1$. If all solenoids are off, then all the quads should be rotated back to normal orientation. And in case of switching off completely all elements of the insertion, only the transformation matrix of the horizontal oscillations will be changed. At this condition, the tune shifts are $\Delta n_x = -0.5$, while $\Delta n_z = -1$. The shift of the working point could be compensated somewhere, say in the interaction region insertion.

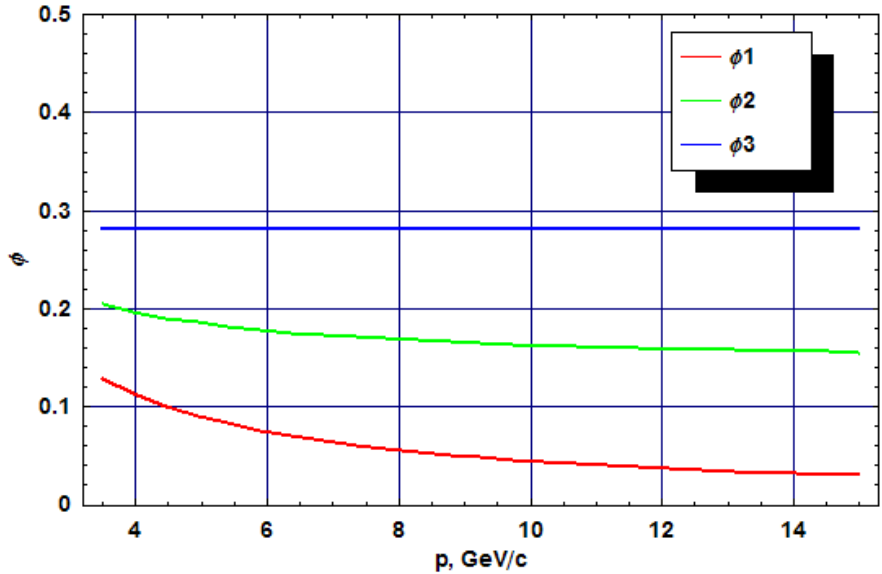

Figure 15. Skew-quads angles vs. the beam energy

## 6.5. Orbit bypasses in long straight

To make the orbit shifts in the long straight section which are necessary to free space for side-by-side housing of two detector systems, PANDA and SEAP, we suggest the "bridge" structure, presented in the Figure 16. The orbit shift is made by two regular bending magnets, identical to those in the HESR arcs. The polarity of the second magnet is negative. The optical functions and dispersion of the bridge are matched with the arcs.

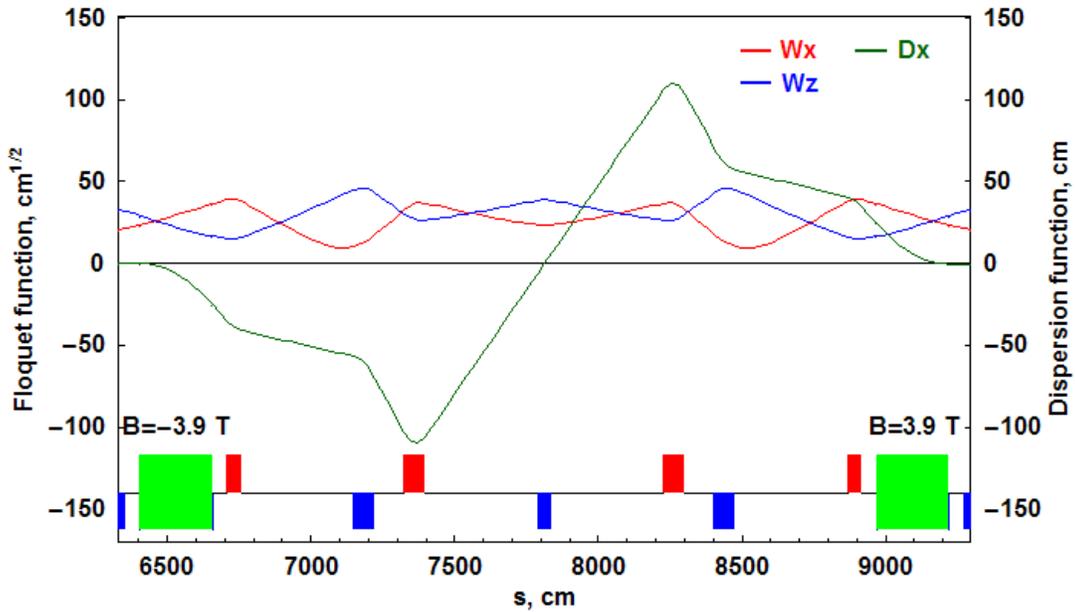

Figure 16. Optical functions of the bypass "bridge"

## 6.6. Interaction region layout

The final focusing in the IP is performed by a quadrupole triplet, which provides $\beta_x = \beta_z$. The optical functions of the interaction region are shown in Figure 17.

The number of bunches $n = 12$ entails the requirement to separate the colliding beams orbit at locations of all parasitic crossings. We suggest here to implement a so-called helical electrostatic orbit separation. For this, the closed orbit distortions in x and z-directions are excited by horizontal and vertical electric fields (30 kV/cm × 100 cm) in the points "a" and "b" respectively. A difference of the x-z phase advance between "a" and "b" is equal to $\pi/2$. Orbits plots with the needed electrostatic separation are shown in the Figure 18. These helical orbits provide the beam separation of at least $\pm 10\sigma_{x,z}$ in all parasitic collision points.

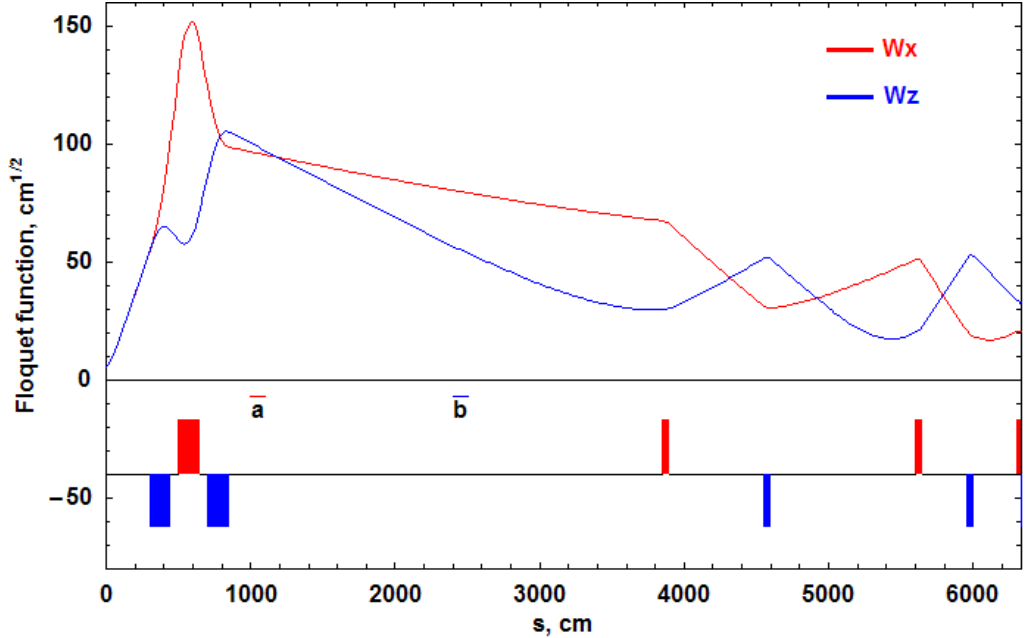

**Figure 17.** Optical functions of the interaction region ($\beta^*_{x,z} = 30$cm)

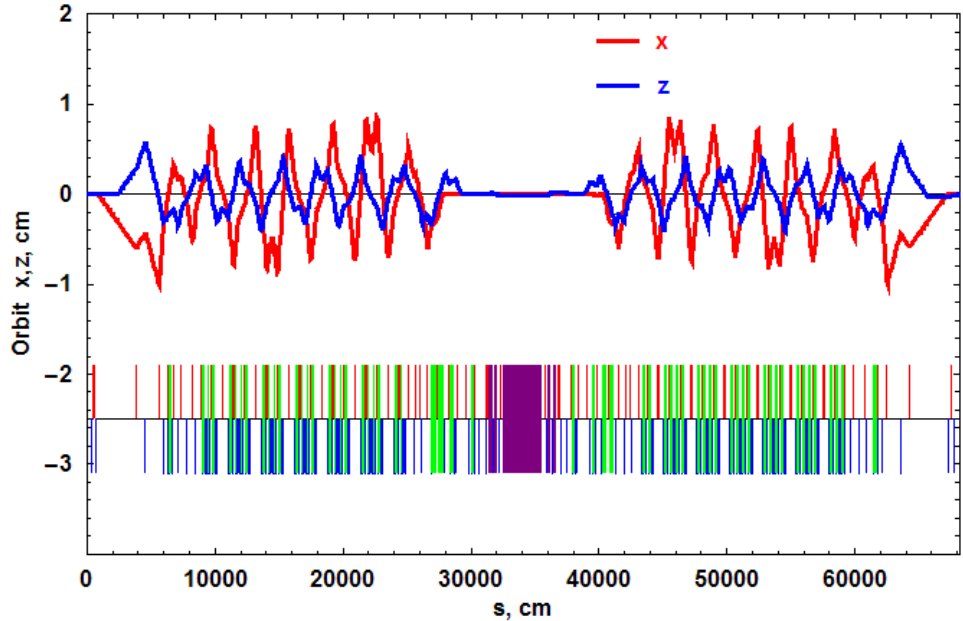
**Figure 18. Electrostatic helical orbit separation**

In the "technical" straight the separation helix is suppressed in the horizontal direction by a small correction of the main separator, and then in the vertical plane by a small steering magnet. So, in the cooler/snake insertion the protons and antiprotons have no deviation from the central closed orbits.

## 7. Asymmetric collider option (3.5 + 15 GeV)

The PAX collaboration [3] has already proposed an asymmetric collider (HESR + modified COSY storage ring), where the beam separation after IP is obtained just exploiting the proton and antiproton energy difference. That proposal did not discuss any scenario of beam operation. Unfortunately, it seems that simultaneous usage of one ring (*COSY*) as booster and part of collider is not possible due to the specific design of the interaction region common for both rings and necessary for beam manipulations.

Another disadvantage of the PAX proposal is the collider option choice with antiprotons circulating in three bunches in the big ring against one proton bunch in the smaller one. In this case the maximal luminosity does not exceed $L = 3 \cdot 10^{30}$ cm$^{-2}$s$^{-1}$. Since there is an evident lack of polarized antiprotons and, on the other hand, polarized proton sources are much more intense, the best way to reach high luminosities is to put antiprotons in the smaller ring for more frequent collisions with higher-energy protons.

We propose to castle the beams: antiprotons are kept at the energy of 3.5 GeV and protons are at 15 GeV as it is shown in the Figure 19. This simple swap helps to reach 3 times more luminosity with 3×10 bunches. This option has one more advantage, it can achieve more dense beams by using coasting beams. In this case the same electron cooling can provide a smaller emittance (down to the space charge limit $e = 0.00003$) for 80 times more intense 15 GeV proton beam.

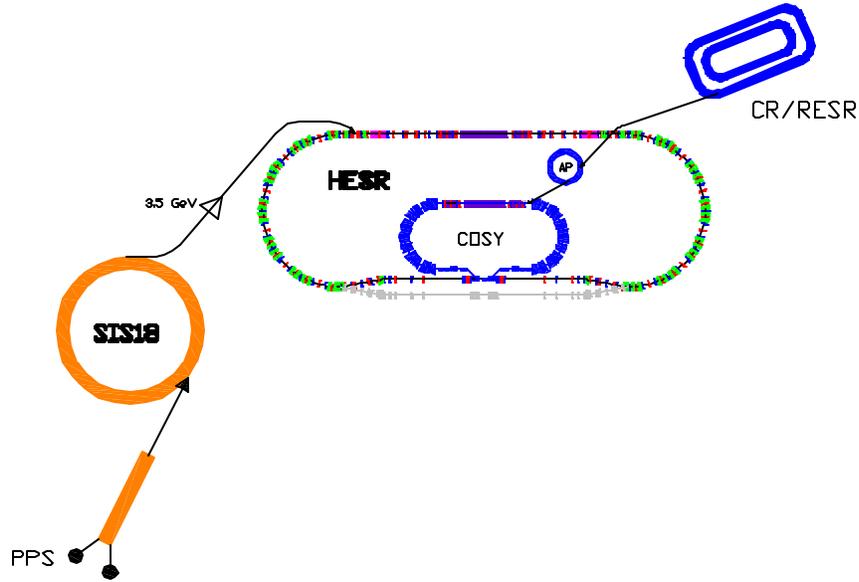

**Figure 19. Scheme of asymmetric collider**

We assume in this collider option, that modified *COSY* serves for accumulation of polarized antiprotons from the AP ring, ramping to an intermediate energy $E_{COSY}=1.207$ GeV for the proton injection in the HESR ($E_{HESR}=$ 3.5 GeV) and then ramping synchronously with HESR up to the experiment energy (3.5 + 15 GeV).

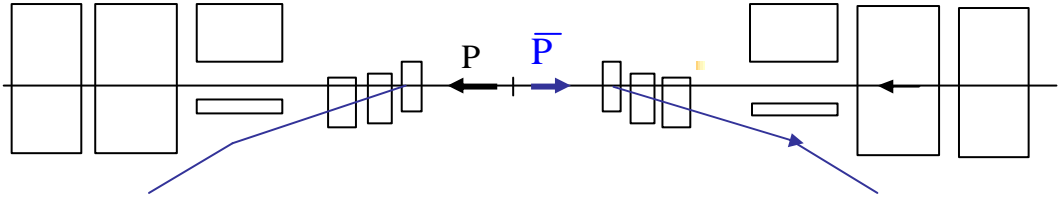

**Figure 20. Schematic drawing of the interaction area (top view)**

Figure 20 presents a final focusing scheme that is to provide a beta-function of 10-30 cm. Superconducting quadrupole magnets of smaller ring are common for both beams. The proton focusing triplet can be made with conventional magnets. The first magnet from the IP is the so called septum qudrupole.

To reach the luminosity limit $L_{max} = 5\cdot10^{31}$ cm$^{-2}$s$^{-1}$ in the discussed configuration, it is easy to calculate that we need to have $N_{\bar{p}}=0.5\times10^{12}$ and $N_p=5\times10^{12}$ with beam emittances $e_1 = e_2 = 0.6\cdot10^{-5}$ cm. At the same time the space charge limit for the coasting beam $\Delta \nu_{1,2} = \dfrac{N_{1,2} r_p}{4\pi \varepsilon_{1,2} \gamma(\gamma^2 -1)}$ is equal 0.05 for the antiproton beam and 0.006 for the proton one.

Both beam-beam parameters for the coasting beams are negligible.

One disadvantage of the coasting colliding beams is a relatively long area of the beam interaction. In Figure 21 a longitudinal distribution of the luminosity is shown for three cases: coasting beams with $\beta_z^* = 30$cm and $\beta_x^* = 100$cm (red); $\beta_z^* = 10$cm and $\beta_x^* = 30$cm (blue) and, for comparison, (dashed line) a cross-section of bunched beam ($\sigma_s = \beta_0 = 30$cm).

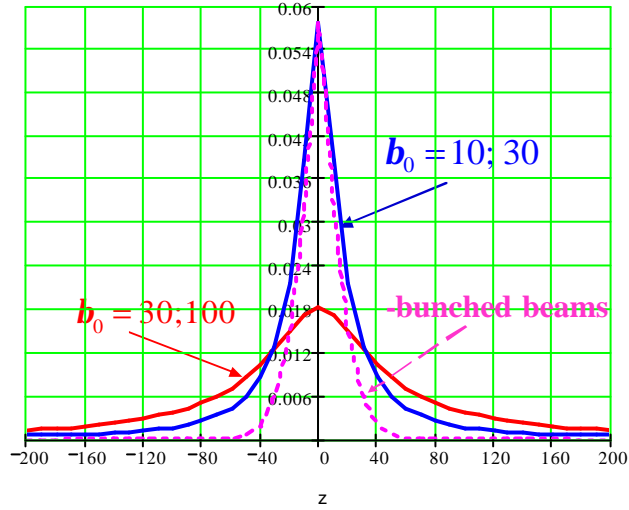

**Figure 21. Longitudinal distribution of the luminosity**

So, one can see that strong final focusing of the coasting beams creates practically the same conditions for detecting events from the collision area as in the case of round bunched beams. At the same time, the limitations from the space charge ($\Delta \nu_1 = 0.1$) and beam-beam interaction ($\xi_1 = 0.02$) restricted the bunched beam luminosity on the level $L = 5 \cdot 10^{30}$ cm$^{-2}$s$^{-1}$. To reach this luminosity $10^{11}$ antiprotons distributed in 10 bunches have to collide at the IP
($\sigma_s = \beta_0 = 30$cm) with 30 proton bunches of total particle intensity $10^{12}$.

## 8. Polarized proton acceleration at SIS-18

Polarized protons will be produced by Polarized Proton Source (protons or negative hydrogen atoms H). After acceleration in the linac the proton beam will be injected into the synchrotron SIS-18. Time of the acceleration from injection up to the top energy of E=6 GeV is about 0.1 sec. Magnetic system of this accelerator consists of 12 periods. The lattice of one SIS-18 period and optical functions are shown in the Figure 22.

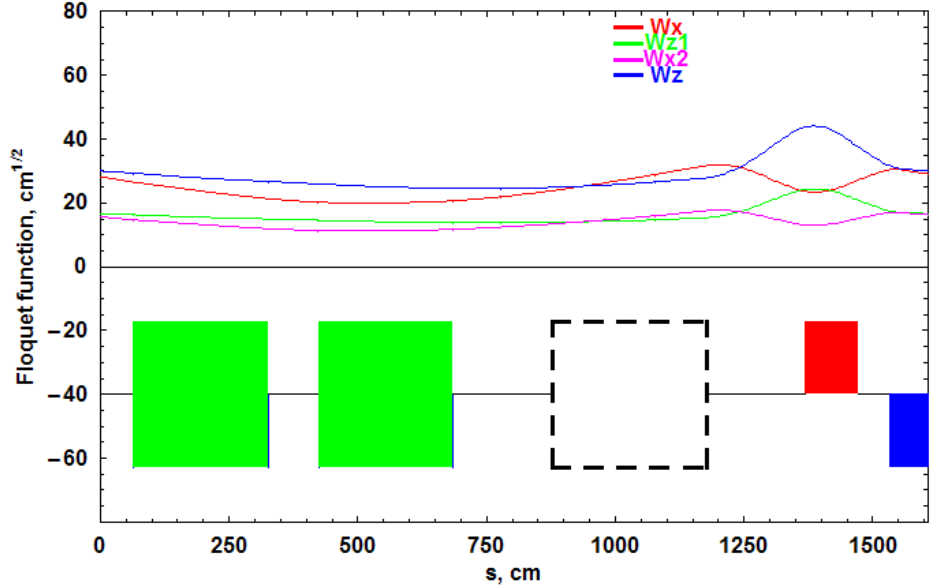

**Figure 22. One period of the SIS-18 lattice**

Let's estimate for SIS-18 the strengths of possible linear spin resonances: $\mathbf{n} = \mathbf{n}_k = k \pm \mathbf{n}_z$. Since the vertical betatron tune is $\mathbf{n}_z = 3.28$, only two linear intrinsic resonances (see Figure 23) are possible in the SIS-18 energy range: $\mathbf{n} = \mathbf{n}_z$ and $\mathbf{n} = 12 - \mathbf{n}_z$ (12 is the machine periodicity). We assume the normalized vertical emittance of proton beam $\mathbf{e}_z = 20\mathbf{p}$ mm·mrad. Then the strengths of these resonances are equal approximately to $w_k(k=0) \approx w_k(k=12) \approx 0.003$ (in units of the Larmor frequency $\mathbf{w}_0$).

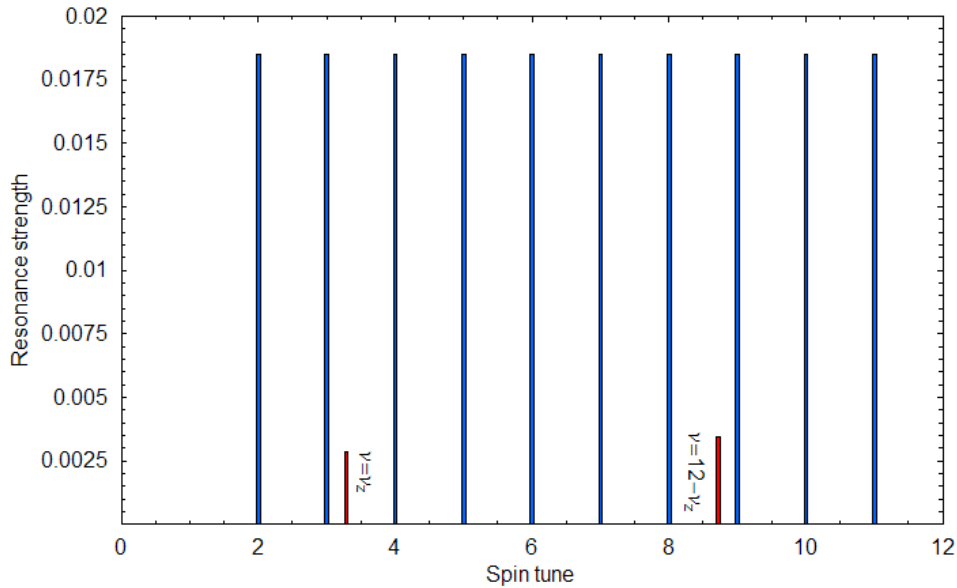

**Figure 23. Strength of linear spin resonances at SIS-18 with partial Siberian snake**

It is well known [8] that the polarization degree after the resonance crossing with the rate $\dot{\boldsymbol{d}} = d\boldsymbol{n}/dt$ is given by the Froissart-Stora formula [8]: $\boldsymbol{z} = \boldsymbol{z}_0 \left(2e^{-y} - 1\right)$, where $y = \dfrac{p w_k^2}{2\dot{d}}$ is the spin phase advance in a resonance zone (tuning $\boldsymbol{d} = \boldsymbol{n} - \boldsymbol{n}_k$ is about $w_k$). A "critical value" of the resonance strength $w_k$ takes place, when $y \approx 1$. With the SIS ramping rate we found: $w_k \approx 0.003$. So, we have to pay more attention to the intrinsic resonances: either to work with a few times smaller beam emittance, or to apply pulsed quads to perform a fast resonance crossing ($\psi \ll 1$).

To suppress possible imperfection resonances ($\boldsymbol{n} = k$) we suggest to introduce a partial Siberian snake into the machine lattice [9]. This snake is a 3 m long solenoid with the field up to 0.5T, which has to be installed in one period of the SIS-18 lattice (see the dashed-line box in the Figure 22). The imperfection resonances excited by the snake are plotted in Fig. 23 (blue lines). The strengths of all imperfection resonances are equal to 0.0185. Such a value of the resonance harmonics $w_k$ is enough to provide, at the ramping rate of 20 T/sec, the adiabatic crossing of every resonance. In these considerations the synchrotron modulation of the proton energy has been also taken into account. It is necessary to remark that the polarization will be reversed by each crossing with exponentially small depolarization ($\dfrac{d\boldsymbol{z}}{\boldsymbol{z}_0} \approx 2e^{-y}$).

A parasitic betatron coupling caused by the snake solenoid gives a shift of the betatron tunes $\Delta \boldsymbol{n} \approx \pm 0.01$.

## 9. Polarized antiprotons

A more serious problem in PAX and this proposal is obtaining of polarized antiprotons. In the Heidelberg experiment a principal possibility to polarize antiprotons by the multi-scattering at polarized hydrogen internal target have been demonstrated [10]. Unfortunately, these data is not enough today for a choice of parameters of the special storage ring for antiproton polarization, aiming to reach as high as possible polarization degree with an optimal intensity of the beam. A new study of the process is needed. One place where it can be done is the Antiproton Deceleration ring (AD) at CERN. A moderate modification of the AD for such experiments has to include polarized target, beam polarimeter, an insertion with a low value of the beta-function and Siberian snake to study the transverse and longitudinal spin component contributions to the final polarization. Measurements should be done at several energies for better understanding of the optimal conditions for antiproton polarization in the AP ring.

Beside this crucial problem we don't see serious troubles to deliver to the collider the antiprotons polarized in an optimized AP ring. In both cases (symmetric and asymmetric options) the COSY type ring has to be equipped with a Siberian snake, which could provide safe conditions for the polarization from injection up to the top energy of 3.5 GeV. A snake design, similar to the discussed above for HESR, is more suitable for this machine including the electron cooler for the asymmetric mode of collider. However, e-cooling at BESSY is not required for the collision inside the HESR, because antiprotons will be cooled in the AP and HESR.

## 10. Cost estimation

Let's make a rough cost estimate of the additional equipments which will be needed to get polarized proton-antiproton colliding beams (Table 3).

**Table 3 Cost estimations of additional equipments**

| Equipment | Cost (M$) Symmetric | Cost (M$) Asymmetric |
|---|---|---|
| Polarized proton source (H⁻) | 1 | 1 |
| Partial Siberian snake + pulse quad for SIS-18 | 0.1 | 0.1 |
| Antiproton Polarization | 10 | 10 |
| Siberian snakes | 2×1=2 | 1+0.3=1.3 |
| Electron cooling (*COSY*) | 3 | 3 |
| IP + bypass | 0.5 | 0.7 |
| Beam separation | 0.5 | — |
| Σ= | 17.1 | 16.1 |

## 11. Summary of the feasibility study

We considered two options of a proton-antipron collider at HESR. The symmetric collider (15 + 15 GeV) can reach the limit luminosity $L_{max} = 5 \cdot 10^{31}$ cm$^{-2}$s$^{-1}$ around the top energy and provide a possibility to carry out experiments in a wide CM energy range, 10–30 GeV. The modifications of the present HESR project needed to arrange collisions of 12 proton bunches with 12 antiproton bunches are relatively moderate.

Practically the same efforts are required to arrange the asymmetric collider. Of course, for this option the maximum achievable energy is two times smaller than that of the symmetric option. But the two-ring collider scheme is capable to collide coasting beams and due to this fact, to get the same limit luminosity level $L_{max} = 5 \cdot 10^{31}$ cm$^{-2}$s$^{-1}$ at the CM energy of 15 GeV.

A more serious problem is clearly the production of polarized antiproton beams. The method proposed by the PAX Collaboration has to be tested with stored antiprotons (for instance at the CERN AD) before being able to finalize the parameters of the AP.

According to our cost estimations, expenses for specific equipments are equal in both variants. Total money is relatively small for the international community, but the collider will open the way for interesting studies, that are not accessible in any other place around the world.

## 12. Acknowledgements

This study has been suggested by the ASSIA Collaboration, and we would like to thank R. Bertini for many useful discussions and for his continuous encouragement. This work has been supported by MIUR (Italian Ministry for Education, University and Research) and INFN (National Institute for Nuclear Physics) through the research program PRIN2003 (Measurement of the Transversity of the Nucleon).

We thank A. Lehrach, B. Lorentz, Yu. Senichev from IKP (Juelich) for kind collaboration in HESR lattice design.

## 13. References

1. FAIR project, http://www.gsi.de/fair/
2. ASSIA LoI, "A study of spin-dependent interactions with antiprotons: the structure of the nucleon", V. Abazov *et al.*, Jan 2004, hep-ex/0507077.
3. PAX Technical proposal, "Antiproton-proton scattering experiments with polarization", Vincenzo Barone *et al.*,. May 2005, hep-ex/0505054.
4. PANDA Collaboration, Technical Progress Report for PANDA, submitted to the QCD-PAC of GSI-FAIR, Jan 2005.
5. Particle Data Group, S. Eidelman *et al.*, Physics Letters B **592**, 1 (2004).
6. V.Parkhomchuk et al., "Proposal of the e-cooling for HESR"
7. Yu.Senichev et al., Proc. of 9-th EPAC, (2004) 653
8. M.Froisart and R.Stora, *Nucl. Instrum. and Methods* **7** (1960), 297
9. Ya.Derbenev and A.Kondratenko, Part.Acc. **8** (1978) 115
10. F.Rathmann et al., Phys. Rev. Lett **71** (1993) 1379